\documentclass[preprint,12pt]{elsarticle}

\usepackage[utf8]{inputenc}
\UseRawInputEncoding
\usepackage{graphicx}
\usepackage{subfigure}
\usepackage{amssymb}
\usepackage{amsthm}
\usepackage{amsfonts}
\usepackage{color}
\usepackage{booktabs}
\usepackage{commath}
\usepackage{cases}
\usepackage{dsfont}

\journal{Computer Physics Communications}
\title{Uncertainty quantification through Monte Carlo method in a cloud computing setting}

\author[a]{Americo Cunha Jr}
\author[b]{Rafael Nasser}
\author[a]{\linebreak Rubens Sampaio}
\author[b]{H\'{e}lio Lopes\corref{1}}
\cortext[1]{Corresponding author. Tel.:+55-21-3527-1500; e-mail: lopes@inf.puc-rio.br}
\author[b]{Karin Breitman}

\address[a]{Department of Mechanical Engineering, PUC--Rio\\
Rua Marqu\^{e}s de S\~{a}o Vicente, 225, G\'{a}vea, Rio de Janeiro - RJ, Brazil - 22453-900 \vspace{5mm}}
\address[b]{Department of Informatics, PUC--Rio\\
Rua Marqu\^{e}s de S\~{a}o Vicente, 225, G\'{a}vea, Rio de Janeiro - RJ, Brazil - 22453-900}

\everymath={\displaystyle}

\renewcommand{\vec}[1]{\ensuremath{\mathbf{#1}}}

\newcommand{\mtx}[1]{\ensuremath{\left[#1\right]}}

\newcommand{\bigO}[1]{\ensuremath{\mathcal{O}( #1 )}}

\newcommand{\indfunc}[1]{\ensuremath{\mathds{1}_{#1} }}

\newcommand{\R}{\ensuremath{\mathbb{R}}}           
\renewcommand{\SS}[1][]{\ensuremath{\Theta^{#1} }}

\newcommand{\SSpt}[1][]{\ensuremath{\theta}}

\newcommand{\randvar}[1]{\ensuremath{#1}}

\newcommand{\randproc}[1]{\ensuremath{#1}}

\newcommand{\pdf}[1]{\ensuremath{p_{\tiny{#1}}}}

\newcommand{\entropy}[1]{\ensuremath{\mathbb{S}\left[  #1 \right] }}

\newcommand{\entropyop}[3]{\ensuremath{ - \int_{#2}^{#3} #1(\xi) \ln \left( #1(\xi) \right) d\xi}}

\newcommand{\expval}[1]{\ensuremath{\mathbb{E}\left[  #1 \right] }}

\begin{document}

\begin{frontmatter}

\begin{abstract}
The Monte Carlo (MC) method is the most common technique used for 
uncertainty quantification, due to its simplicity and good statistical results. 
However, its  computational cost is extremely high, and, in many cases, prohibitive. 
Fortunately, the MC algorithm is easily parallelizable, which allows its use in 
simulations  where the computation of a single realization is very costly. 
This work presents a methodology for the parallelization of the MC method, 
in the context of cloud computing. This strategy is based on the MapReduce 
paradigm, and allows an efficient distribution of tasks in the cloud. This methodology 
is illustrated on a problem of structural dynamics that is subject to uncertainties. 
The results show that the technique is capable of producing good results concerning 
statistical moments of low order. It is shown that even a simple problem may require 
many realizations for convergence of histograms, which makes the cloud computing 
strategy very attractive (due to its high scalability capacity and low-cost). Additionally, 
the results regarding the time of processing and storage space usage  allow one to qualify 
this new methodology as a solution for simulations that require a number of 
MC realizations beyond the standard.
\end{abstract}

\end{frontmatter}

\section{Introduction}

Most of the predictions that are necessary for decision making
in engineering, economics, actuarial sciences, and so on., are made based 
on computer models. These models are based on assumptions 
that may or may not be in accordance with reality. Thus, a model can have uncertainties 
on its predictions, due to possible wrong assumptions made during its conception. This source of 
variability on the response of a model is called \emph{model uncertainty} \cite{soize2005p1333}. 
In addition to modeling errors, the response of a model
is also subject to variabilities due to uncertainties on model parameters, 
which may be due to measurement errors, imperfections in the manufacturing 
process, and other factors. This second source of randomness on the models response is 
called \emph{data uncertainty} \cite{soize2005p1333}.

One way to take into account these uncertainties is to use the 
theory of probability, to describe the uncertain parameters as
random variables, random processes, and/or random fields.
This approach allows one to obtain a model where it is possible
to quantify the variability of the response. For instance, the
reader can see from \cite{ritto2012p111,ritto2013p145} where 
techniques of stochastic modeling are applied to describe
the dynamics of a drillstring. Other applications in
structural dynamics can be seen in \cite{ritto2009p865,ritto2011p373}.
It is also worth mentioning the contributions of \cite{zio2012p145},
in the context of hydraulic fracturing, \cite{lopes2012p39}, for estimation 
of financial reserves, and in \cite{clement2013p61} for the analysis of 
structures built by heterogeneous hyperelastic materials.
For a deeper insight into stochastic modeling, with an emphasis
in structural dynamics, the reader is encouraged to read 
\cite{soize2005p1333,soize2005p623,soize2013p2379}.

\begin{figure}[h!]
	\centering
	\includegraphics[scale=0.5]{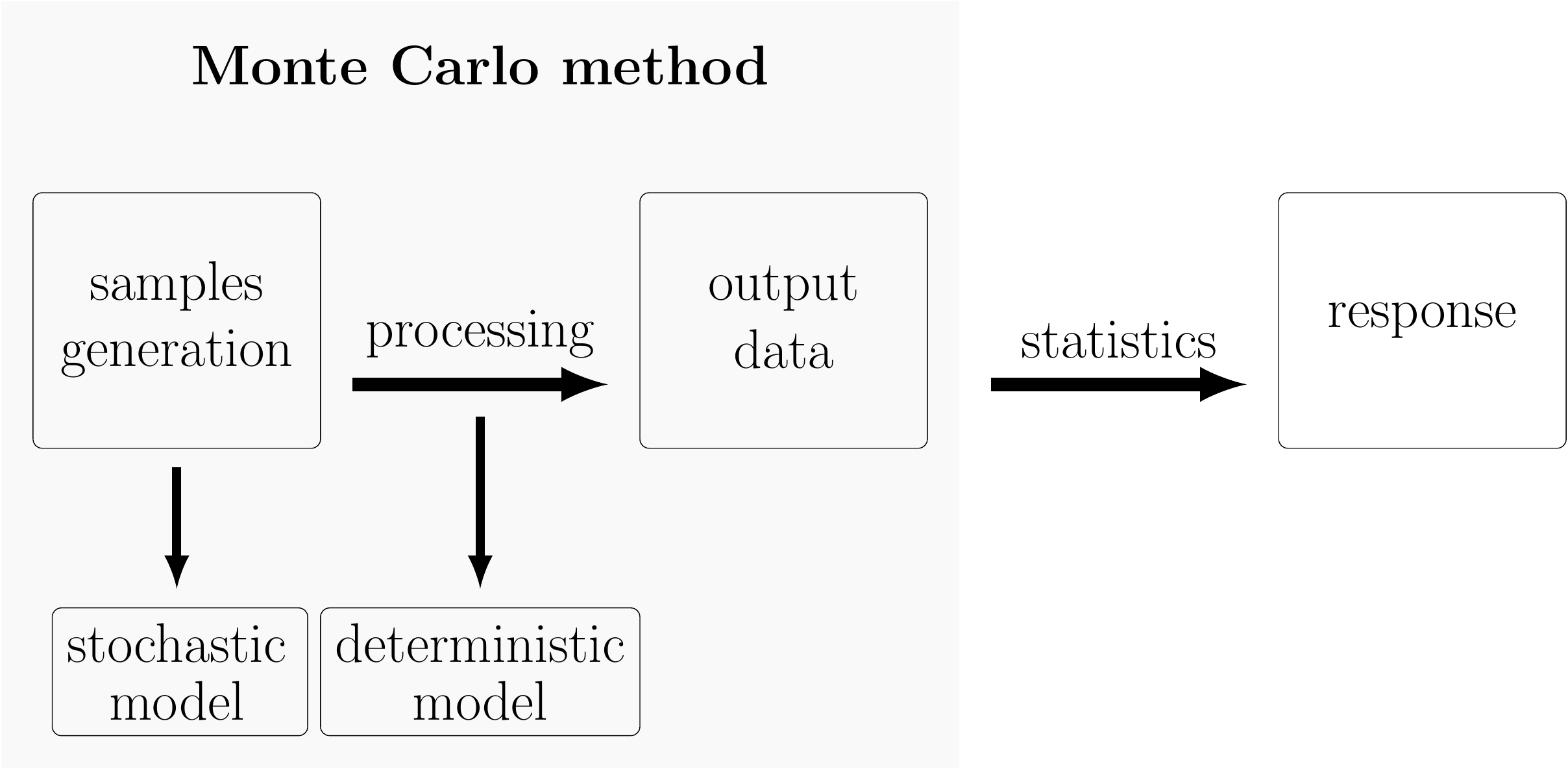}
	\caption{General overview of Monte Carlo algorithm.}
	\label{mc_method_fig}
\end{figure}

To compute the propagation of uncertainties of the random parameters 
through the model, the most used technique in the literature is the 
Monte Carlo (MC) method \cite{metropolis1949p335}.
This technique generates several realizations (samples) of the random parameters 
according to their distributions (stochastic model). Each of these realizations
defines a deterministic problem, which is solved (processing) using a deterministic
technique, generating an amount of data. Then, all of these data are combined through 
statistics to access the response of the random system under analysis
\cite{liu2001,shonkwiler2009,casella2010}. A general overview of the MC algorithm
can be seen in the Figure~\ref{mc_method_fig}.

The MC method does not require that one implements a new computer code 
to simulate a stochastic model. If a deterministic code to simulate a similar 
deterministic model is available, the stochastic simulation can be performed 
by running the deterministic program several times, changing only the 
parameters that are randomly generated.
This nonintrusive characteristic is a great advantage of MC when 
compared with other methods for uncertainty quantification, such as 
generalized Polynomial~Chaos~(gPC), \cite{xiu2002p619}, 
which demands a new code for each new random system that one wants to simulate.
Additionally, if the MC simulation is performed for a large number of samples, it completely 
describes the statistical behavior of the random system. 
Unfortunately, MC is a very time-consuming method, which makes unfeasible 
its use for complex simulations, when the  processing time of a single realization 
is very large or the number of realizations to an accurate result is huge
\cite{liu2001,shonkwiler2009,casella2010}.

Meanwhile, the MC method algorithm can easily be parallelized because each realization 
can be done separately and then aggregated to calculate the statistics. 
The parallelization of the MC algorithm allows one to obtain significant gains in terms of 
processing time, which can enable the use of the MC method in complex simulations.
In this context, cloud computing can be a very fruitful tool to enable the use of the MC method 
to access complex stochastic models because it is a natural environment for implementation 
of parallelization strategies. Moreover, in theory, cloud computing offers almost infinite scalability 
in terms of storage space, memory and processing capability, with a financial cost significantly 
lower than the one that is necessary to acquire a traditional cluster with the same capacity.

In this spirit, this work presents a methodology for implementing the MC method
in a cloud computing setting, which is inspired by the
MapReduce paradigm \cite{dean2004}. This approach consists of splitting,
among several instances of the cloud environment, the MC calculation, 
processing each one of these tasks in parallel, and finally merging the results into 
a single instance to compute the statistics. As an example, the methodology is 
applied to a simple problem of stochastic structural dynamics.
The use of cloud in not new in the context of engineering and sciences 
\cite{Shiers2009}. We would like to mention the work of Ari and Muhtaroglu 
\cite{Ari2012} that proposes a cloud computing service for finite element analysis, 
the work of Jorissen et al. \cite{Jorissen2012} that proposes a scientific cloud 
computing platform that offers high performance computation capability for 
materials simulations, and the work of Wang et al. \cite{Wang2011} that discusses 
the Cumulus cloud based project with its applications to scientific computing, 
just to cite a few.

This paper is organized as follows. Section~\ref{cloud_comp} makes a brief 
presentation of the cloud computing concept. Section~\ref{parallel_mc} presents
a parallelization strategy for the MC method in the context of cloud computing.
Section~\ref{case_study} describes the case of study in which the proposed 
methodology is exemplified. Section~\ref{num_experim} presents and discusses 
the statistics done with the data and the convergence of the results.
Finally, section~\ref{concl_remarks} presents the conclusions and highlights 
the main contribution of this work.


\section{Cloud computing}
\label{cloud_comp}

Traditionally, the term cloud is a metaphor about the way the Internet is usually 
represented in network diagrams. In these diagrams, the icon of the cloud represents 
all the technologies that make the Internet work, ignoring the infrastructure and 
complexity that it includes. Likewise, the term cloud has been used as an abstraction 
for a combination of various computer technologies and paradigms, e.g., virtualization, 
utility computing, grid computing, service-oriented architecture and others, 
which together provide computational resources on demand, such as storage, 
database, bandwidth and processing \cite{velte2009}.

Therefore, cloud computing can be understood as a style of computing where 
information technology capabilities are elastic, scalable, and are provided 
as services to the users via the Internet \cite{cearley2009,vaquero2008p50}.
In this style of computing, the computational resources are provided for the users
on demand, as a pay-as-you-go business model, where they only need to pay
for the resources that were effectively used. Due to its great potential for solving 
practical problems of computing, it is recognized as one of the top five 
emerging technologies that will have a major impact on the quality of science 
and society over the next 20 years \cite{buyya2010}.

The reader can see from \cite{roloff2012p371} a detailed comparison between 
three cloud providers (Amazon EC2, Microsoft Azure and Rackspace) and a 
traditional cluster of machines. These experiments were done using the well-known 
NAS parallel benchmarks as an example of general scientific application. That article 
demonstrates that the cloud can have a higher performance and cost efficiency than 
a traditional cluster. 

Furthermore, a traditional cluster require huge investments in hardware and 
in their maintenance and one can not ``turn off resources contracts", 
while they are unnecessary, to save money. Then, traditional clusters are almost 
prohibitive for scientific research without large financial resources. 

In a cloud computing environment one pays only per hour of use of one virtual machine. 
Other costs of this platform are the shared/redundant storage and data transfers. 
For the data transfer, all inbound data transfers (i.e., data going to the cloud) 
are free and the price for outbound data transfers (i.e., data going out of the cloud) 
is a small cost that depends on the volume of data transferred.

Given these characteristics, it is easy to imagine a situation where computational resources 
can be turned on and off according to demand, providing  unprecedented savings compared 
with acquisition and maintenance of a traditional cluster. In addition, if it is possible to 
parallelize the execution, the total duration of the process can be minimized using 
more virtual machines of the cloud.


\section{Parallelization of Monte Carlo method in the cloud}
\label{parallel_mc}

The strategy to run the MC algorithm in parallel, as proposed in this work 
is influenced by the MapReduce paradigm \cite{dean2004}, which was originally 
presented to support the processing of large collections of data in parallel 
and distributed environments.
This paradigm consists in two phases: the first (Map) divides the computational 
job into several partitions and each partition is executed in parallel by different 
machines; the second phase (Reduce) collects the partial results returned by each 
machine, aggregates partial results and computes a response to the computational job.

\begin{figure}[h!]
	\centering
	\includegraphics[scale=0.26]{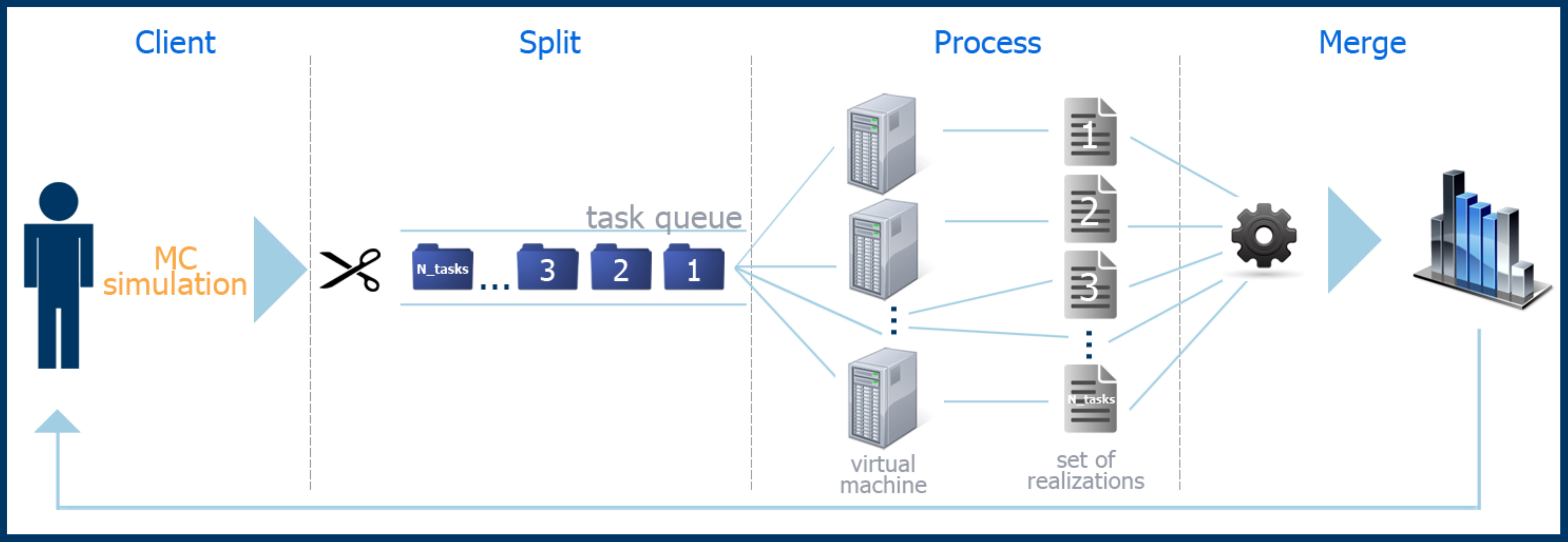}
	\caption{General overview of Monte Carlo parallelization strategy in the cloud.}
	\label{parallel_mc_fig}
\end{figure}

We propose a MapReduce strategy for parallel execution in the cloud of the MC method 
that is composed of three steps: \emph{split}, \emph{process}, and \emph{merge}. 
The split and the process steps correspond to the Map, while the merge corresponds to the Reduce step.
This strategy of parallelization was implemented in a cloud computing setting 
called McCloud \cite{nasser2012,nasser2013}, which runs on the Microsoft Windows Azure 
platform (http://www.windowsazure.com). A general overview of the strategy 
can be seen in Figure~\ref{parallel_mc_fig}, and a detailed description of each step 
is made below.

\subsection{Split}
\label{split}

First of all, the split step establishes the number of cloud virtual machines to be used and turns then on.
Then, it divides a MC simulation with $N_{MC}$ realizations into tasks and puts them into a queue.
Each one of these tasks is composed of an ensemble of $N_{serial}$ realizations to be simulated. 
Thus, it is necessary to process a number of tasks equal to $N_{tasks}=N_{MC}/N_{serial}$. 
These tasks are distributed in a uniform manner (approximately)  among the virtual machines 
\cite{nasser2012,nasser2013}. It is important to note that the number of tasks and virtual machines influences 
directly in the total simulation processing time and the financial expenses of the cloud computing service.

In the simulations above, the realizations of the random parameters are obtained by the use of a
pseudorandom numbers generator. This generator deterministically constructs sequences of numbers indexed by 
the value of a  seed, which emulates a set of random numbers. 
It happens that, if two instances of tasks have the same seed, they will 
generate the same sequence of random numbers. In this case, part of these results will 
be redundant at the end of the simulation. 

To avoid the possibility of repeated seeds, it is necessary to adopt a strategy of seed distribution 
among the virtual machines. This strategy must generate one seed for each virtual machine,
and guarantee that the sequence of random numbers generated in each one of these machines 
is different from the sequences generated on the other machines. There are several ways to 
define a distribution strategy and we will not discuss this in detail because it is quite simple. 
To see the strategy adopted in the example of section~\ref{num_experim}, the reader may consult
\cite{nasser2012,nasser2013}.

\subsection{Process}
\label{process}

The process step uses the available virtual machines to pick and process, in an 
asynchronous way, task by task in the queue. Thus, in theory, the execution is 
as fast as the number of virtual machines used.
In practice, there is a limit to efficiency gain because of the existence of overhead, 
such as input/output of data and managing parallelization. Thus, one of the 
challenges that must be solved for an efficient simulation, at a low-cost, is the 
determination of an optimal number of virtual machines to be used 
\cite{nasser2012,nasser2013}.
Currently, this optimization process occurs empirically. However, in future works,
we will seek to rationalize it, according to the nature of the simulations involved.

At this step, there is a criterion of tolerance against failures. When a task is caught
from the queue by a virtual machine, it has a time limit to be executed. 
If it is not performed in this time, it is put back into the queue for another virtual machine 
to try and run it. Thus, a hardware or machine communication problem does not 
overtake the execution of MC simulation.

The output data generated by each executed task is saved onto the hard disk for 
subsequent post-processing. The total amount of storage space used by a MC 
simulation is proportional to the number of realizations. Therefore, this step 
also requires attention in terms of storage space usage because the demand for 
hard disk space may become unfeasible for a simulation with a large number of 
realizations.

\begin{figure}[h!]
	\centering
	\includegraphics[scale=0.61]{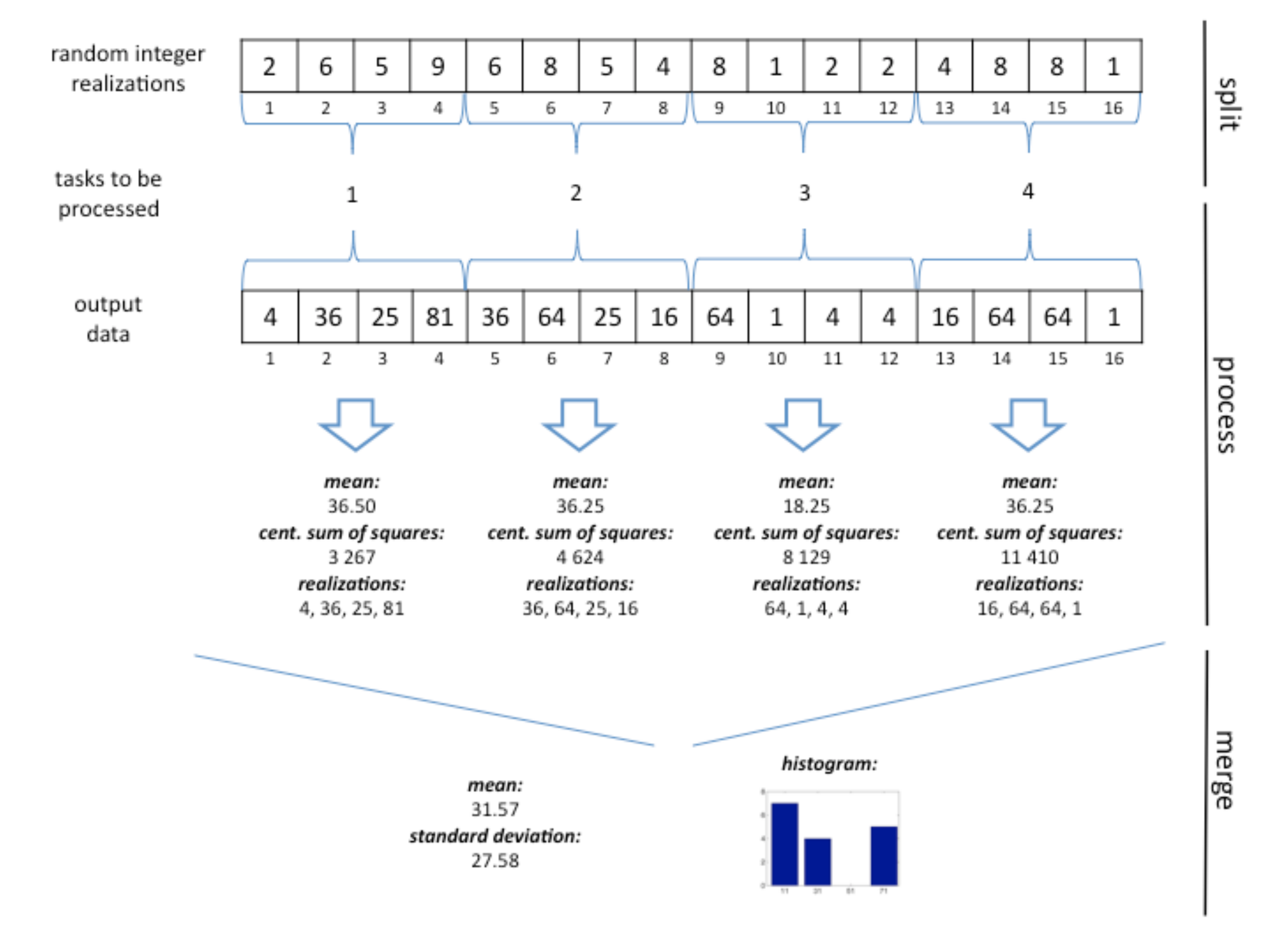}
	\caption{Exemplification of the strategy for parallel calculation of statistics.}
	\label{example_statistics_fig}
\end{figure}

To reduce the storage space usage in the example of section~\ref{num_experim}, 
we chose to calculate the mean and standard deviation using the strategy of
pairwise parallel and incremental updates of the statistical moments 
described in \cite{bennett2009}, which uses the Welford-Knuth algorithm
\cite{welford1962p419,knuth1998v2}. Thus, for each executed task,
instead of saving all the simulation data for subsequent calculation of
the statistical moments, we save only the mean and the centered sum of squares.
For the calculation of the histograms, the random variables of interest were identified 
before the processing step, and their realizations were saved for being used in the 
histogram construction, during the post-processing step. This strategy is exemplified 
in Figure~\ref{example_statistics_fig}, which shows  the parallelization of a MC simulation,
with 16 realizations,  that aims to calculate the square of an integer random number 
between 0 and 9.

\subsection{Merge}
\label{merge}

The merge step starts when the last task in a virtual machine finishes. This can occur
in any virtual machine. This step reads, from the hard disk, all the information contained
in the output data from the simulations, and combines them through statistics to
obtain relevant information about the problem under analysis.
At the end of this stage, the saved data are discarded and only the merged result 
is stored, to reduce future costs of data storage.


\section{Case of study}
\label{case_study}

\subsection{Physical system}

\begin{figure}[h!]
	\centering
	\includegraphics[scale=0.75]{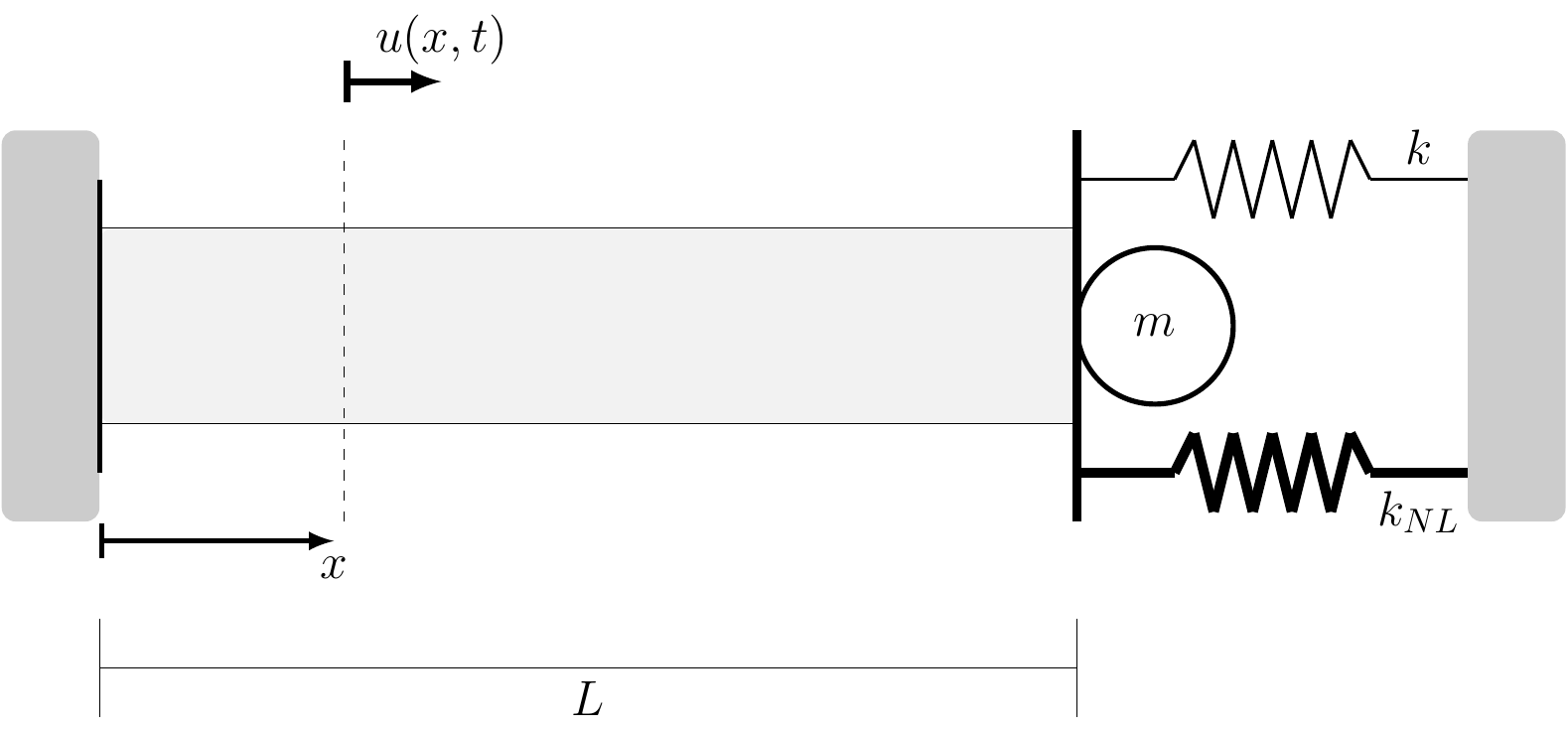}
	\caption{Sketch of a bar fixed at one end and attached
					to two springs and a mass on the other end.}
	\label{bar_fig}
\end{figure}

The system of interest in this study case is an elastic bar fixed at a rigid wall,
on the left side, and attached to a lumped mass and two springs (one linear and
one nonlinear), on the right side, such as illustrated in Figure~\ref{bar_fig}.
The stochastic nonlinear dynamics of this system was investigated in
\cite{cunhajr2012p2673,cunhajr2013a,cunhajr2013b,cunhajr2013sub}, 
where the reader can see more details about the modeling procedure presented below. 
For simplicity, from now on, this system will be called the fixed-mass-spring 
bar or simply the bar.

\subsection{Model equation}

The physical quantity of interest is the bar displacement field $u$,
which depends on the position $x$ and the time $t$, and evolves, 
for all $(x,t) \in (0,L) \times (0,T)$, according 
to the following hyperbolic partial differential equation

\begin{equation}
    \rho A \dpd[2]{u}{t} + c \dpd{u}{t} = 
    \dpd{}{x}\left( E A \dpd{u}{x}  \right)  + f(x,t),
    \label{bar_eq}
\end{equation}

\noindent
where $\rho$ is mass density, $A$ is the cross section area,
$c$ is the damping coefficient, $E$ is the elastic modulus, and
$f$ is a distributed external force, which depends on $x$ and $t$.

The boundary conditions for this problem are given by

\begin{equation}
    u(0,t) = 0,
    \label{bar_bc1}
\end{equation}

\noindent
and

\begin{equation}
    E A \dpd{u}{x} (L,t) = -ku(L,t) -k_{NL}\left( u(L,t) \right)^3 - m \dpd[2]{u}{t}(L,t),
    \label{bar_bc2}
\end{equation}

\noindent
where $k$ is the stiffness of the linear spring, $k_{NL}$ is the stiffness of the nonlinear spring,
and $m$ is the lumped mass.

The initial position and the initial velocity of the bar are

\begin{equation}
    u(x,0) = u_0(x),
    \label{bar_ic1}
\end{equation}

\noindent
and

\begin{equation}
    \dpd{u}{t}(x,0) = v_0(x),
    \label{bar_ic2}
\end{equation}

\noindent
where $u_0$ and $v_0$ are known functions of $x$, defined for $0 \leq x \leq L$.

\subsection{Discretization of the model equation}
\label{galerkin_form}

To approximate the solution of the initial/boundary value problem 
given by Eqs.(\ref{bar_eq}) to (\ref{bar_ic2}), we employ the Galerkin 
method \citep{hughes2000}. This results in the following
system of ordinary differential equations

\begin{equation}
    \mtx{M} \ddot{\vec{u}}(t) + \mtx{C} \dot{\vec{u}}(t) +
    \mtx{K}\vec{u}(t) =  \vec{f}(t) + \vec{f}_{NL} \left( \dot{\vec{u}}(t) \right),
    \label{galerkin_eq}
\end{equation}

\noindent
supplemented by the following pair of initial conditions

\begin{equation}
    \vec{u}(0) =  \vec{u}_0 \qquad \mbox{and} \qquad \vec{\dot{u}}(0) =  \vec{v}_0,
    \label{galerkin_ic_eq}
\end{equation}

\noindent
where $\vec{u}(t)$ is the vector of $\R^N$ in which the $n$-th component 
is the $u_n(t)$, $\mtx{M}$ is the mass matrix, $\mtx{C}$ is the damping matrix, and 
$\mtx{K}$ is the stiffness matrix.  Additionally, $\vec{f}(t)$, $\vec{f}_{NL}\left(\vec{u}(t)\right)$, 
$\vec{u}_0$, and $\vec{v}_0$ are vectors of $\R^N$, which respectively 
represent the external force, the nonlinear force, the initial position, 
and the initial velocity. The initial value problem defined by 
Eqs.(\ref{galerkin_eq}) and (\ref{galerkin_ic_eq}) has its solution approximated
by the Newmark method \citep{newmark1959p67,hughes2000}.

\subsection{Stochastic model}
\label{stochastic_model}

To introduce randomness in the bar model, we assume that
the external force $f$ is a random field proportional to a 
normalized Gaussian white noise. Moreover, the elastic modulus 
is assumed to be a random variable.

The probability distribution of $\randvar{E}$ is characterized by its
probability density function (PDF) $\pdf{\randvar{E}}: (0,\infty) \to \R$,
which is specified, based only on the known information about this 
parameter, by  the maximum entropy principle 
\cite{soize2000p277,shannon1948p379, jaynes1957p620, jaynes1957p171}.

The maximum entropy principle says that, among all the probability 
distributions consistent with the current known information of 
$\randvar{E}$, to choose the one that maximizes its entropy.
Thus, to specify $\pdf{\randvar{E}}$, it is necessary to maximize 
the entropy function 

\begin{equation}
	\entropy{\pdf{\randvar{E}}} = \entropyop{ \pdf{\randvar{E}} }{0}{\infty},
	\label{dif_entropy}
\end{equation}

\noindent
subjected to the constraints (known information) imposed by

\begin{equation}
	\int_{0}^{\infty} \pdf{\randvar{E}}(\xi) d\xi = 1,
	\label{norm_cond_E}
\end{equation}

\begin{equation}
	\expval{\randvar{E}} = \mu_{\randvar{E}} < \infty,
	\label{meanval_E}
\end{equation}

\noindent
and

\begin{equation}
	\expval{\ln \left( \randvar{E} \right) } < \infty,
	\label{finite_explog_E}
\end{equation}

\noindent
where $\expval{\cdot}$ is the expected value operator, and
$\mu_{\randvar{E}}$ is the mean value of $\randvar{E}$.

Regarding the known information, the Eq.(\ref{norm_cond_E}) is 
the normalization condition of the random variable, 
the Eq.(\ref{meanval_E}) means that the mean value of 
$\randvar{E}$ is known, and the Eq.(\ref{finite_explog_E}) is a 
sufficient condition to ensure that $\randproc{U}$ have 
finite variance \cite{soize2000p277}.

The desired distribution is the gamma, whose PDF is given by

\begin{equation}
	\pdf{\randvar{E}}(\xi) = \indfunc{(0,\infty)}
	\frac{1}{\mu_{\randvar{E}}} 
	\left( \frac{1}{\delta_{\randvar{E}}^2} \right)^{ \displaystyle \left( \frac{1}{\delta_{\randvar{E}}^2} \right) }
	\frac{1}{\Gamma(1/\delta_{\randvar{E}}^2)} 
	\left( \frac{\xi}{\mu_{\randvar{E}}} \right)^{ \displaystyle \left( \frac{1}{\delta_{\randvar{E}}^2}-1 \right) }
	\exp\left( - \frac{\xi}{\delta_{\randvar{E}}^2 \mu_{\randvar{E}}} \right),
	\label{gamma_distrib}
\end{equation}

\noindent
where the symbol $\indfunc{(0,\infty)}$ denotes the indicator function of the 
interval $(0,\infty)$, $\delta_{\randvar{E}}$ is a dispersion factor, and 
the $\Gamma$ indicates the gamma function.

Due to the randomness of $\randproc{F}$ and $\randvar{E}$,
the displacement of the bar becomes a random field $\randproc{U}$,
which evolves according to the following stochastic partial differential 
equation

\begin{equation}
    \rho A \dpd[2]{\randproc{U}}{t} + 
    c \dpd{\randproc{U}}{t} = 
    \dpd{}{x} \left( \randvar{E}(\SSpt) A \dpd{\randproc{U}}{x} \right)
    + \randproc{F} (x,t,\SSpt).
    \label{randbar_eq}
\end{equation}

The boundary conditions now read as

\begin{equation}
    \randproc{U}(0,t,\SSpt) = 0,
    \label{randbar_bc1}
\end{equation}

\noindent
and

\begin{equation}
    E A \dpd{\randproc{U}}{x} (L,t,\SSpt) = 
    -k \randproc{U}(L,t,\SSpt) 
    -k_{NL}\left( \randproc{U}(L,t,\SSpt) \right)^3 
    - m \dpd[2]{\randproc{U}}{t}(L,t,\SSpt),
    \label{randbar_bc2}
\end{equation}

\noindent
for $0 < t < T$  and $\SSpt \in \SS$, while the initial conditions are

\begin{equation}
    \randproc{U} (x, 0, \SSpt) = u_0(x),
    \label{randbar_ic1}
\end{equation}

\noindent
and

\begin{equation}
    \dpd{\randproc{U}}{t} (x, 0, \SSpt) = v_0(x),
    \label{randbar_ic2}
\end{equation}

\noindent
for $0 \leq x \leq L$ and $\SSpt \in \SS$, where $\SS$ denotes
the sample space in which we are working.


\section{Numerical experiments}
\label{num_experim}

We employ the MC method in the cloud to approximate the solution of the stochastic 
initial/boundary value problem defined by Eqs.(\ref{randbar_eq}) to (\ref{randbar_ic2}).
This procedure uses a sampling strategy with the number of realizations always 
being equal to a power of four. In this procedure, each realization of the random 
parameters defines a new initial value problem given by Eqs.(\ref{galerkin_eq}) and
(\ref{galerkin_ic_eq}), which is solved deterministically as described in 
section~\ref{galerkin_form}. Then, these results are combined through statistics.

The implementation of the MC method was conducted in MATLAB, with the aid of two
executable files. The first one, which is executed in each one of the virtual machines,
generates realizations of the random parameters and performs the determinist 
calculations to solve the associated variational problem. The other executable 
computes the statistics of the output data generated by the first executable.

\subsection{Probability density function}

A random variable is completely characterized by its PDF. The knowledge 
of the PDF allows us to obtain all the statistical moments of the random variable 
and to calculate the probability of any event associated with it. So, we start our
analysis with the PDF estimation.

\begin{figure} [h!]
				\centering
				\subfigure[$16,384$ realizations]{
				\includegraphics[scale=0.35]{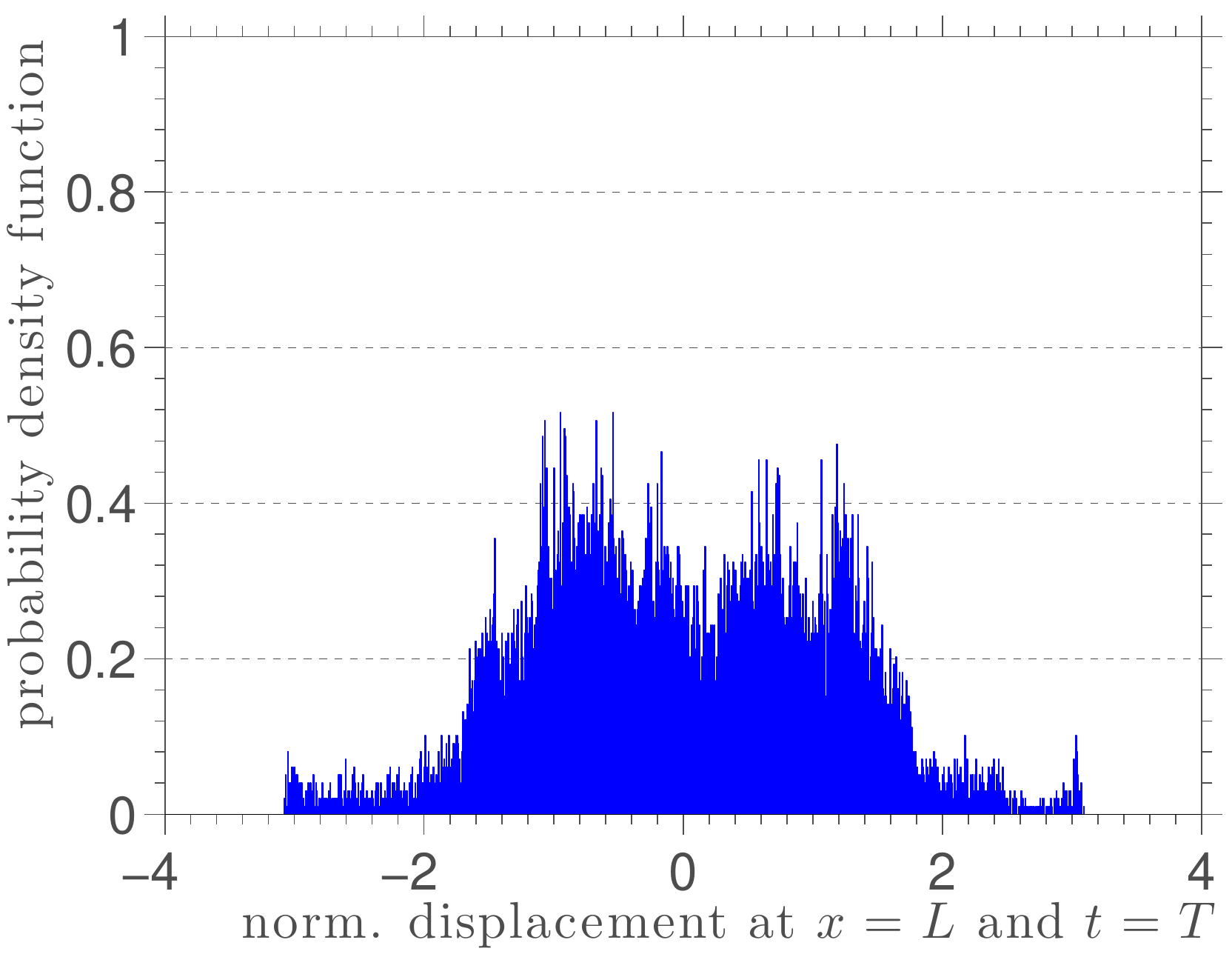}}
				\subfigure[$65,536$ realizations]{
				\includegraphics[scale=0.35]{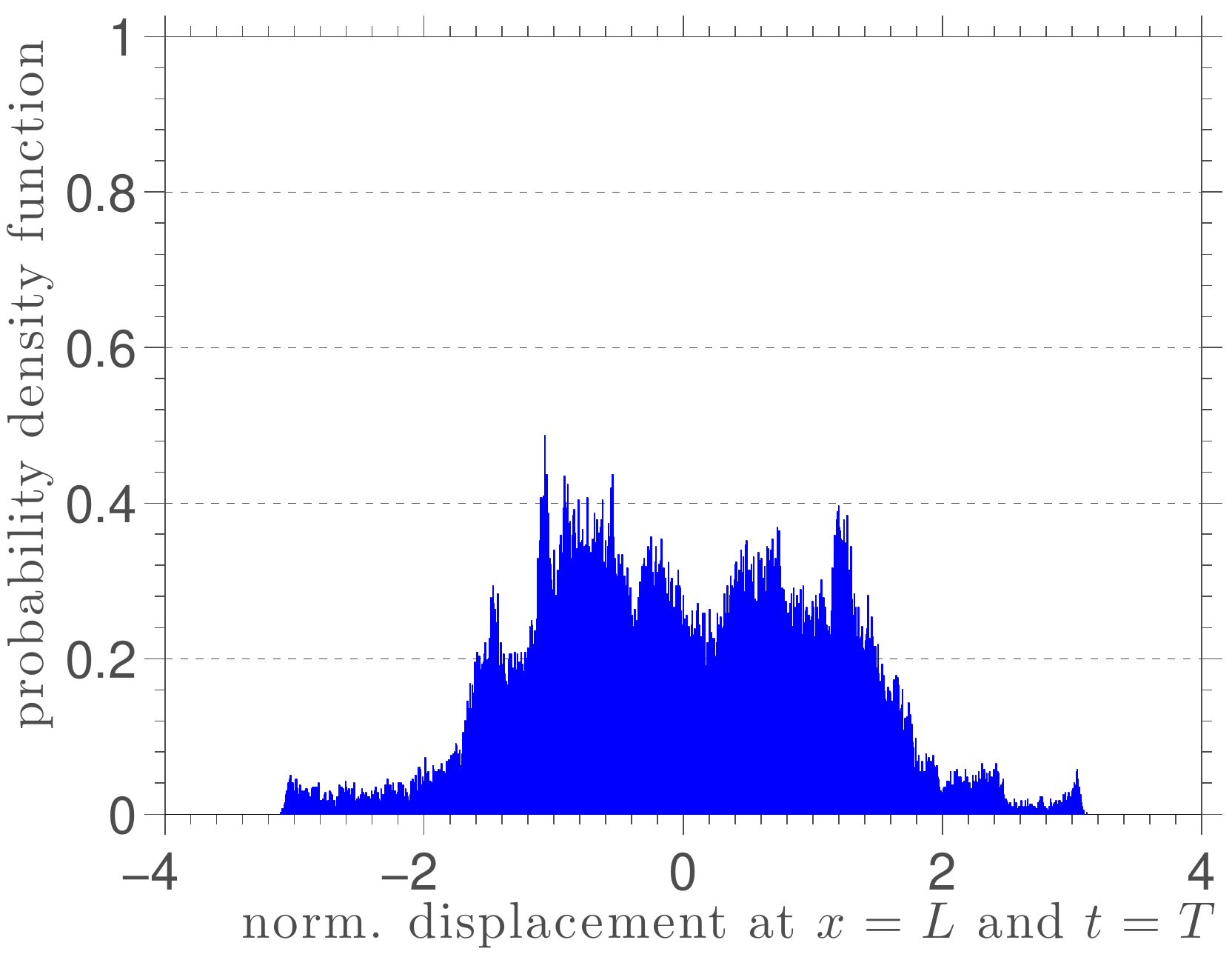}}\\
				\subfigure[$262,144$ realizations]{
				\includegraphics[scale=0.35]{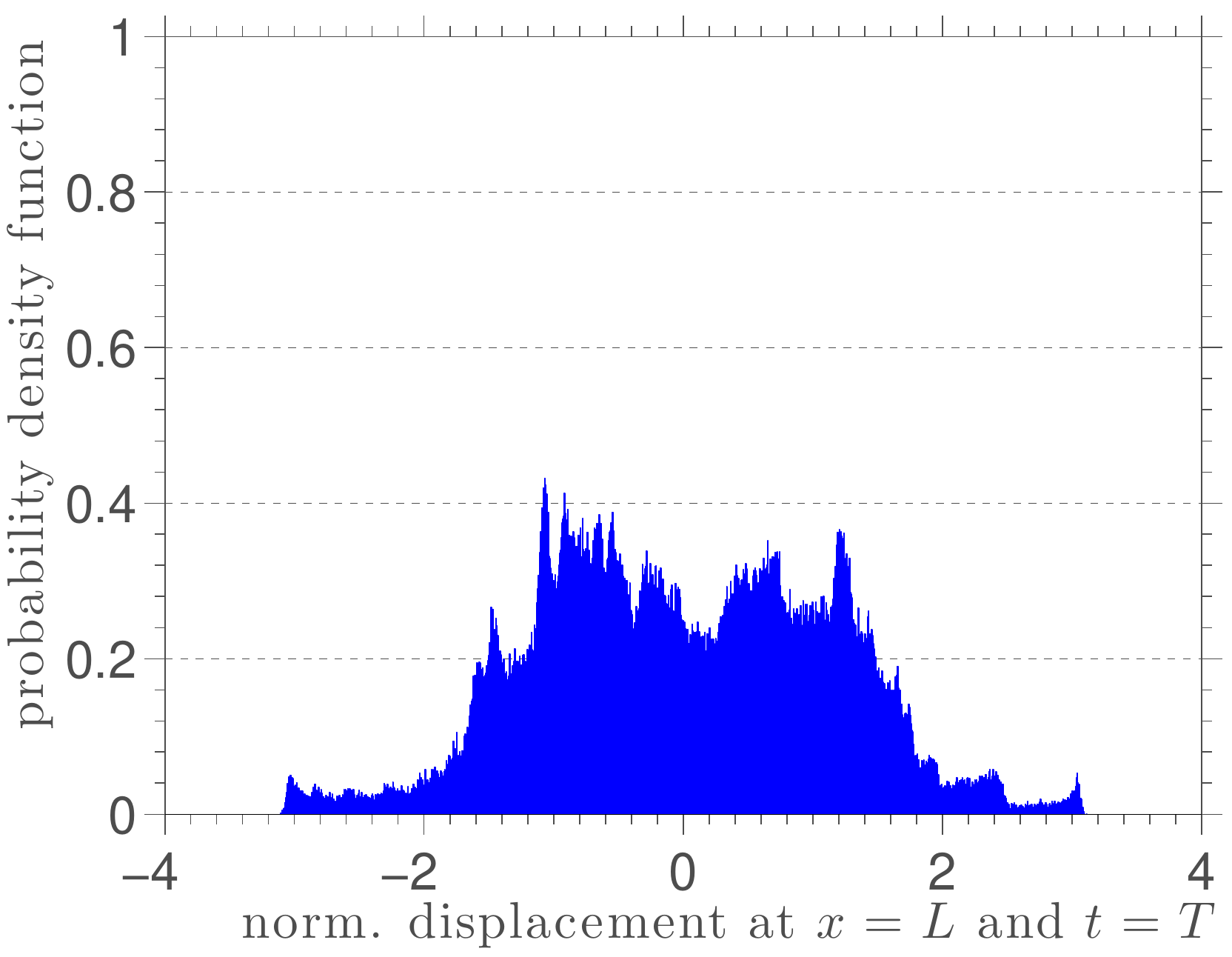}}
				\subfigure[$1,048,576$ realizations]{
				\includegraphics[scale=0.35]{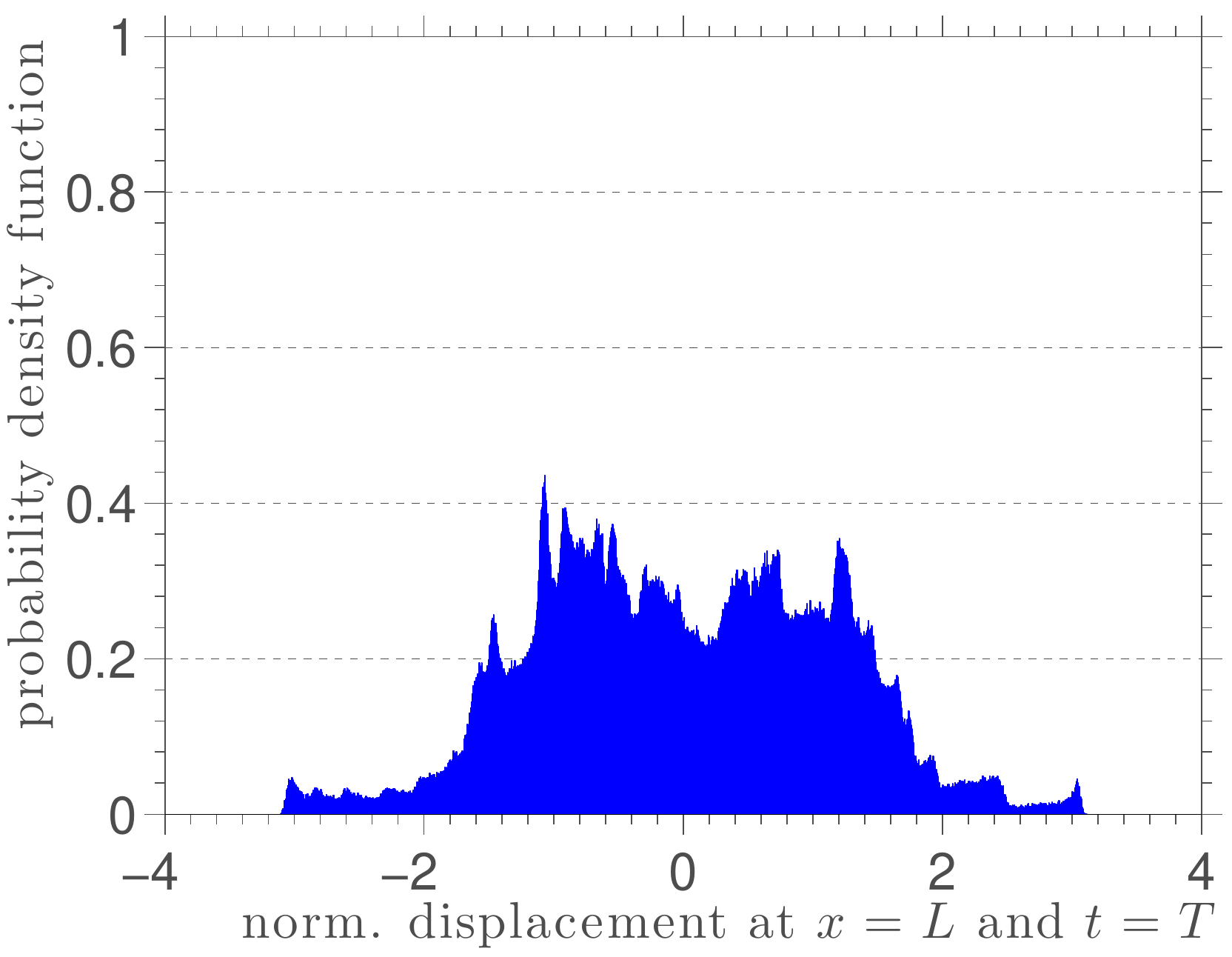}}
				\caption{This Figure illustrates estimations for the PDF of the (normalized)
							random variable $\randproc{U}(L,T,\cdot)$, for different values of 
							the total number of realizations in MC simulations: (a) $16,384$ realizations, (b) $65,536$ realizations, (c) $262,144$ realizations and (d) $1,048,576$ realizations.}
				\label{pdf_uL_fig}
\end{figure}

The estimations for the PDF of the (normalized\footnote{By normalized we mean
a random variable with zero mean and unit standard deviation}) bar right extreme 
displacement for a fixed instant of time, is shown in Figure~\ref{pdf_uL_fig} for 
different values of the total number of realizations in MC simulations. 
We can note that as the number of samples in the MC simulation increases, small 
differences may be noted on the peaks of successive estimations of the PDF. 

We use a convergence criterion based on a residue of the random variable $\randproc{U}(L,T,\SSpt)$,
defined as the absolute value of the difference between two successive approximations of 
$\pdf{\randproc{U}(L,T,\cdot)}$, i.e.,

\begin{equation}
	R_{\randproc{U}(L,T,\cdot)} = 
	\left| \pdf{\randproc{U}(L,T,\cdot)}^{4n} - \pdf{\randproc{U}(L,T,\cdot)}^{n} \right|,
	\label{res_pdf_def}
\end{equation}

\noindent
where the superscript $n$ indicates the number of realizations in the MC simulation.
In this case, we say that the MC simulation reached a satisfactory result if
this residue is less than a prescribed tolerance $\epsilon$, i.e., 
$R_{\randproc{U}(L,T,\cdot)} < \epsilon$ for all $\SSpt \in \SS$.
For instance, $\epsilon = 0.05$.

\begin{figure} [h!]
				\centering
				\includegraphics[scale=0.45]{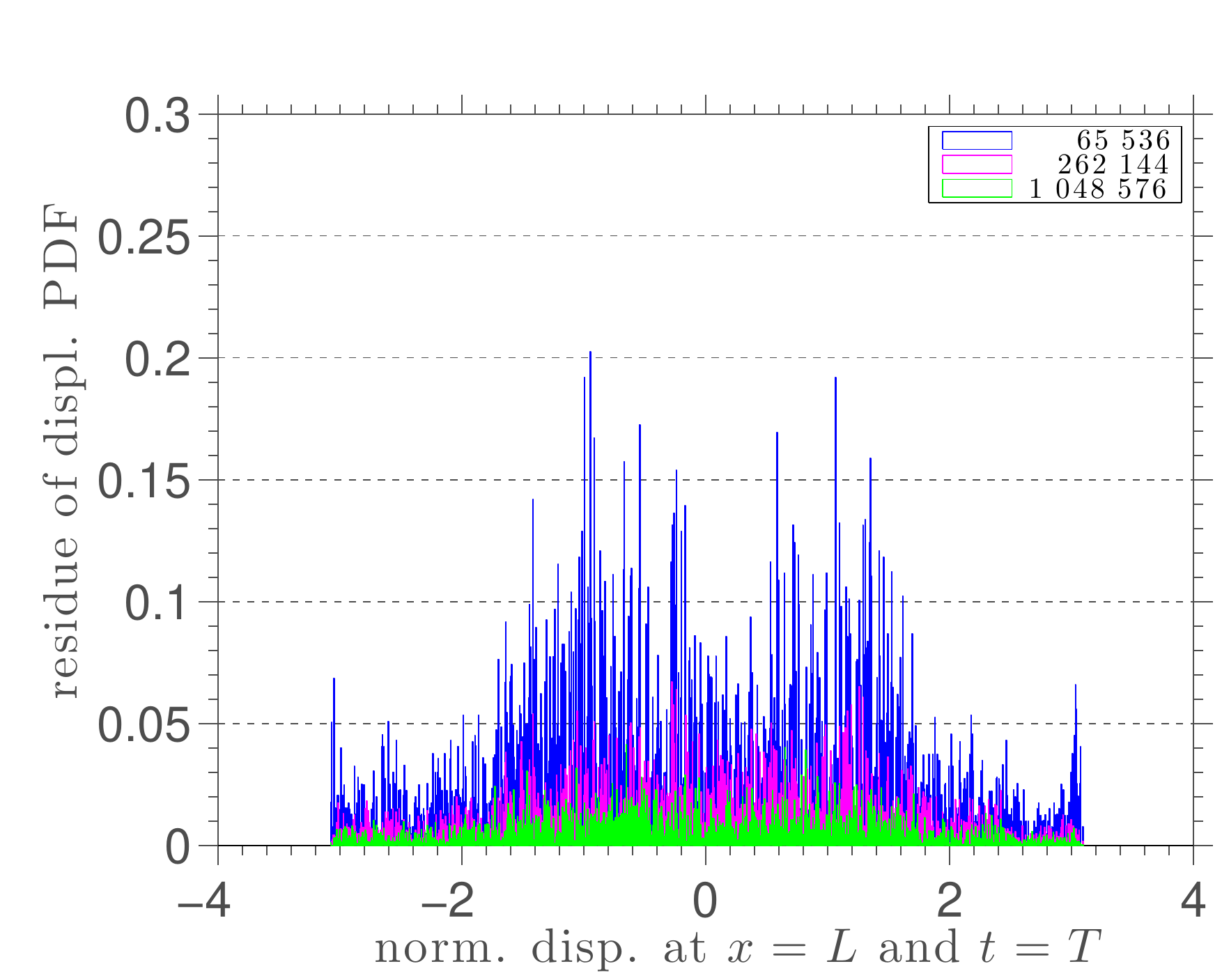}
				\caption{This figure illustrates the residue of the 
							$\randproc{U}(L,T,\cdot)$ PDF.}
				\label{res_uLT_fig}
\end{figure}

The reader can observe the distribution of the residue of $\randproc{U}(L,T,\SSpt)$,
for several values of MC realizations, in the Figure~\ref{res_uLT_fig}. Note that although 
the residue decreases with the increase of the MC realizations, only one simulation with
$1,048,576$ samples was able to fulfill the convergence criterion.

Therefore, despite the number of samples used in the MC simulation is very high and 
the problem is relatively simple, the tolerance achieved was relatively low.
It is common to observe in the literature some works that analyze problems
much more complex, for instance 
\cite{spanos2008p456,liang2011p147,adhikari2012p229}, among many others, 
using some hundreds of samples.

This example leads us to reflect about the number of realizations required 
to obtain statistical independence of the results. In this context, the use of 
MC in a cloud computing setting appears to be
a viable solution, able to make the work feasible at a low-cost.

\subsection{Mean and standard deviation}

\begin{figure} [h!]
				\centering
				\subfigure[$16,384$ realizations]{
				\includegraphics[scale=0.35]{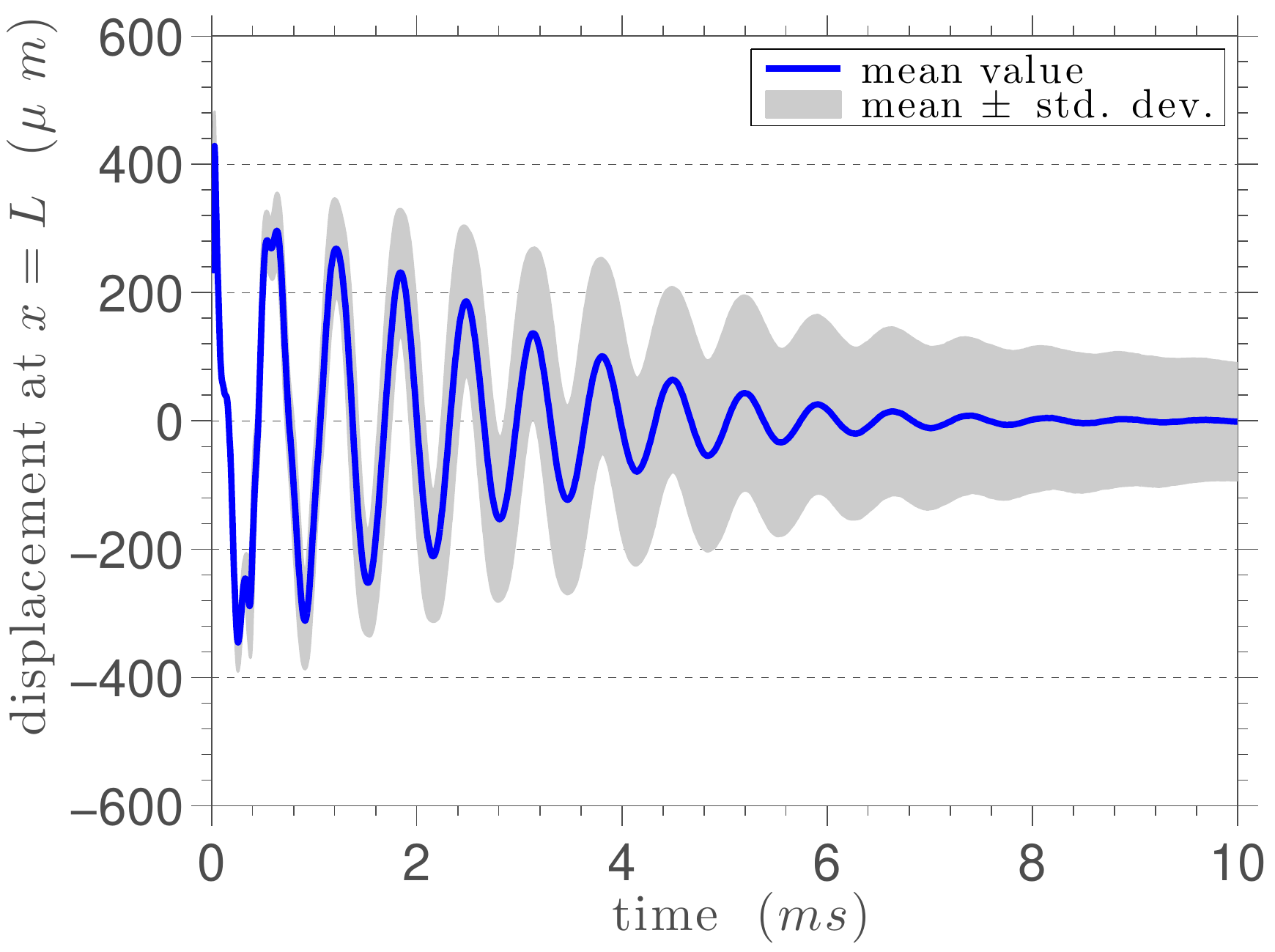}}
				\subfigure[$65,536$ realizations]{
				\includegraphics[scale=0.35]{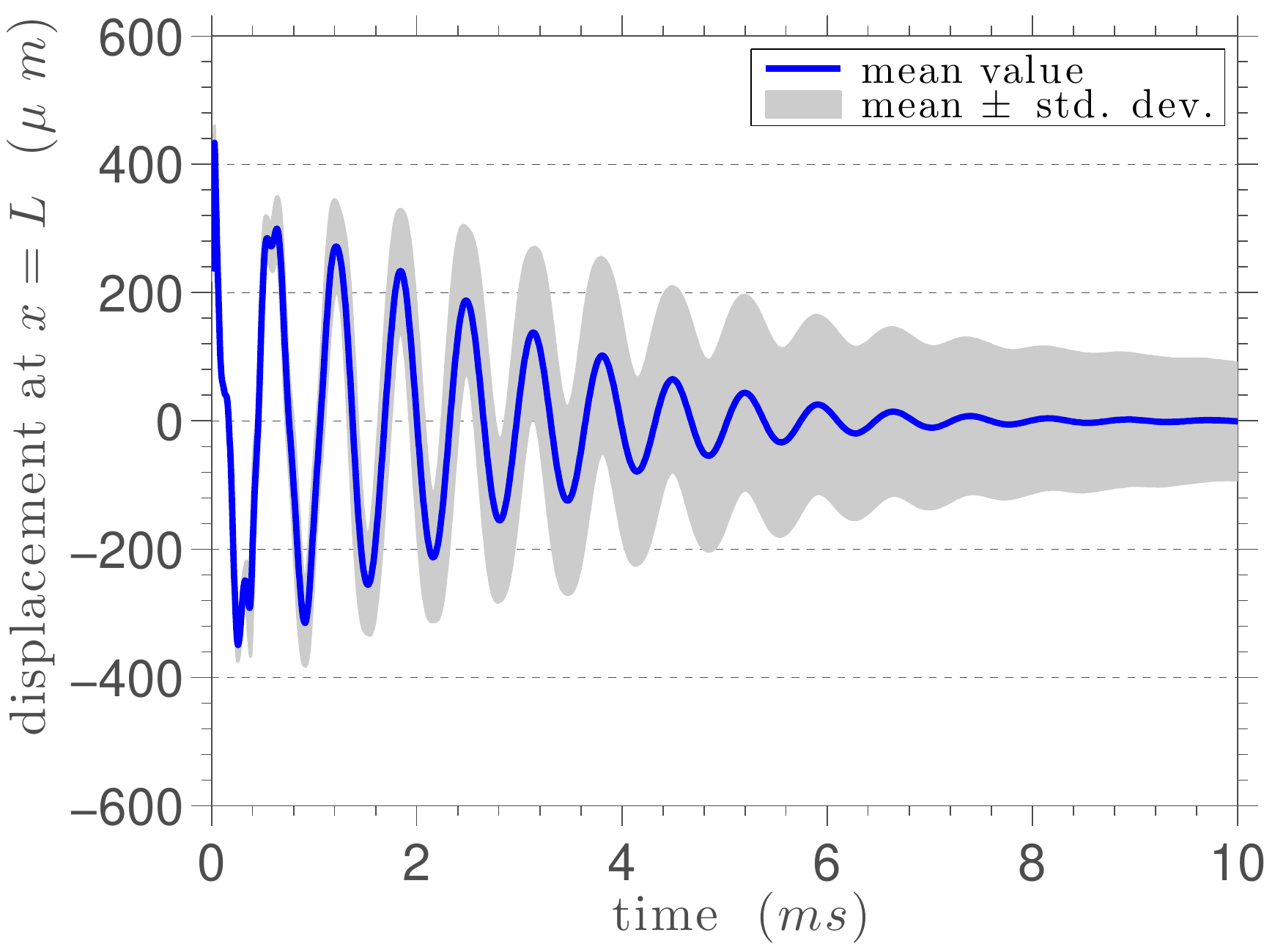}}\\
				\subfigure[$262,144$ realizations]{
				\includegraphics[scale=0.35]{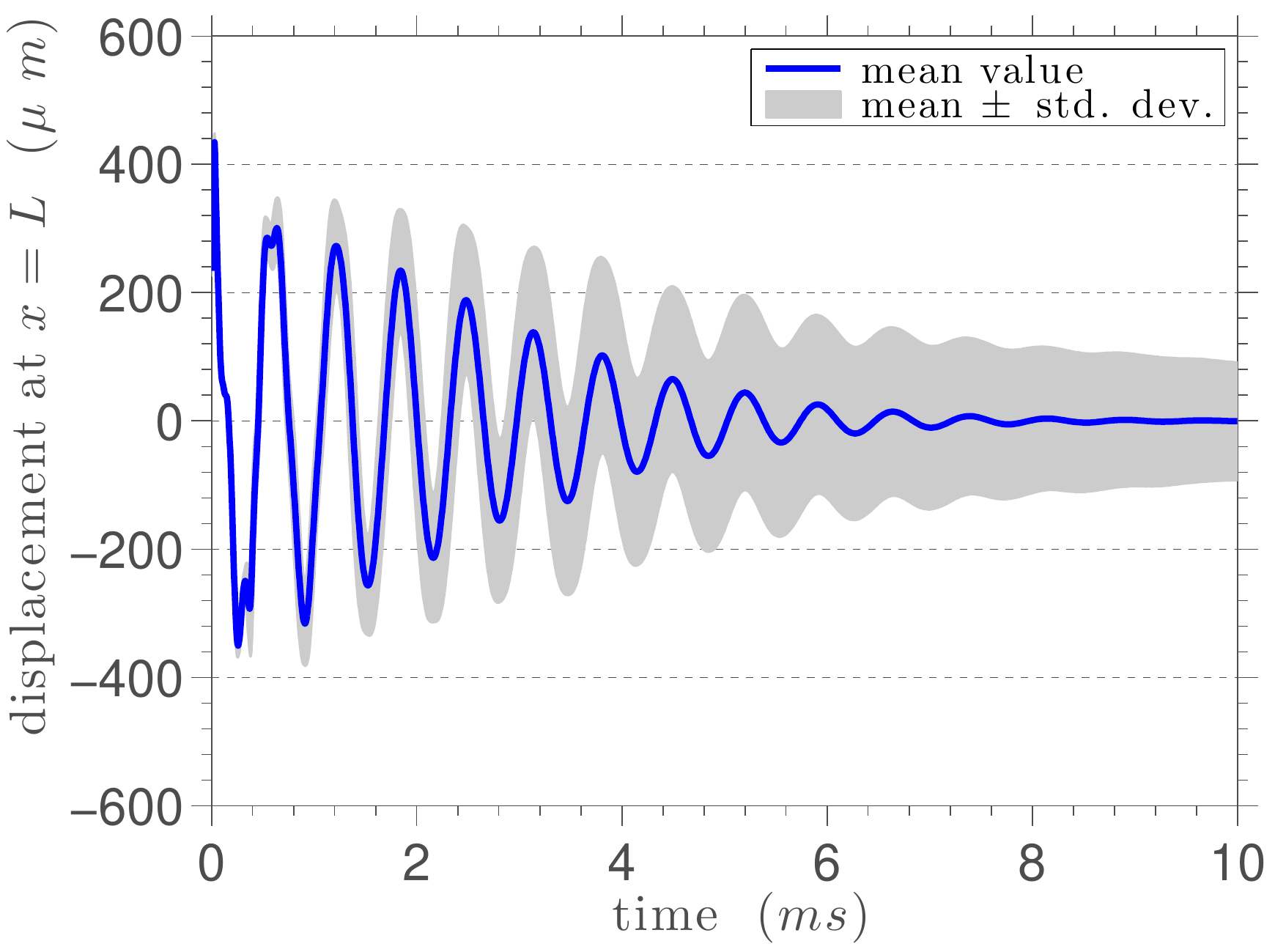}}
				\subfigure[$1,048,576$ realizations]{
				\includegraphics[scale=0.35]{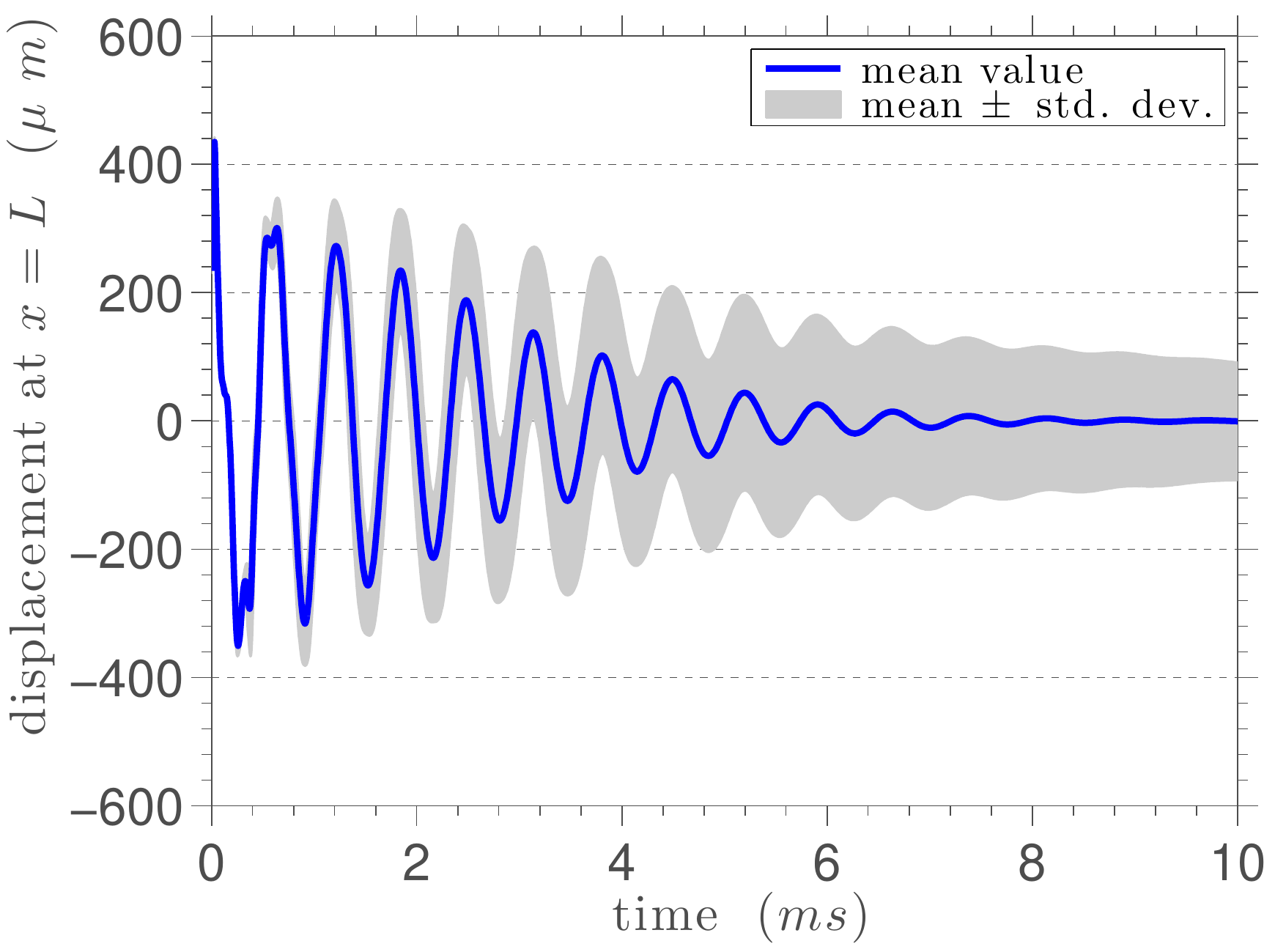}}
				\caption{This Figure illustrates the mean value (blue line) and a confidence interval
							(grey shadow) with one standard deviation, of the random process 
							$\randproc{U}(L,\cdot,\cdot)$, for different values of the total number 
							of realizations in MC simulations: : (a) $16,384$ realizations, (b) $65,536$ realizations, (c) $262,144$ realizations and (d) $1,048,576$ realizations.}
				\label{ci_uL_fig}
\end{figure}

Figure~\ref{ci_uL_fig} shows the evolution of the bar right extreme 
displacement mean (blue line) and an envelope of reliability (grey shadow)
around it, obtained by adding and subtracting one standard deviation 
around the mean. This figure shows these graphs for different values 
of the total number of realizations in MC simulations.

The first conclusion we can draw from these results is that the 
low-order statistics, from the qualitative point of view, do not 
undergo major changes when the total number of realizations 
is higher than $16,384$. However, based on a purely visual analysis,
we can not conclude anything quantitatively.

To investigate the quantitative differences on the results, we define the residue
of $\randproc{U}(L,\cdot,\cdot)$ mean or standard deviation similarly to the one
defined to Eq.(\ref{res_pdf_def}), by changing the PDF for the mean or standard
deviation of $\randproc{U}(L,\cdot,\cdot)$ only.

\begin{figure} [h!]
				\centering
				\includegraphics[scale=0.45]{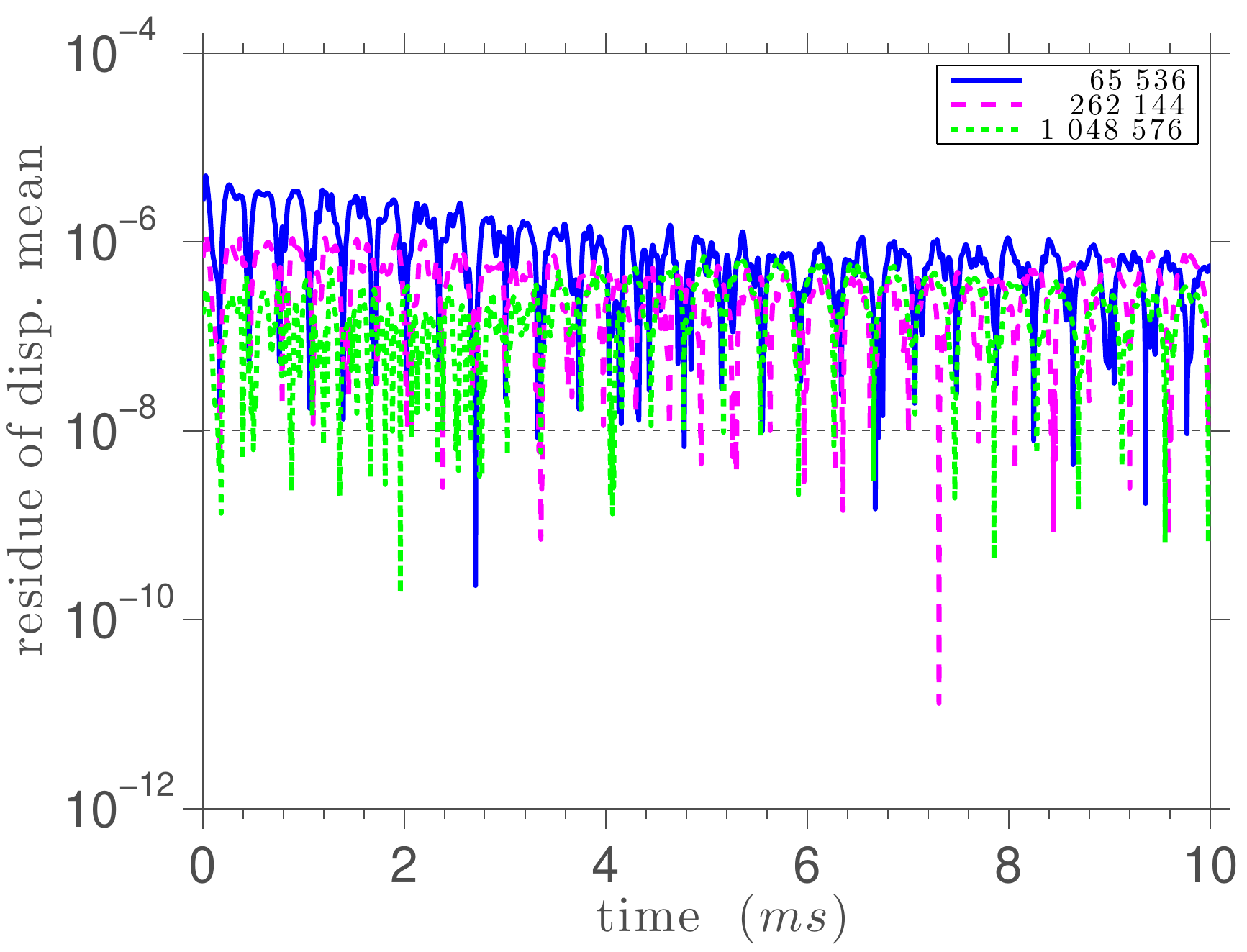}
				\caption{This figure illustrates the evolution of residue of the 
							$\randproc{U}(L,\cdot,\cdot)$ mean.}
				\label{res_mean_uL_fig}
\end{figure}

\begin{figure} [h!]
				\centering
				\includegraphics[scale=0.45]{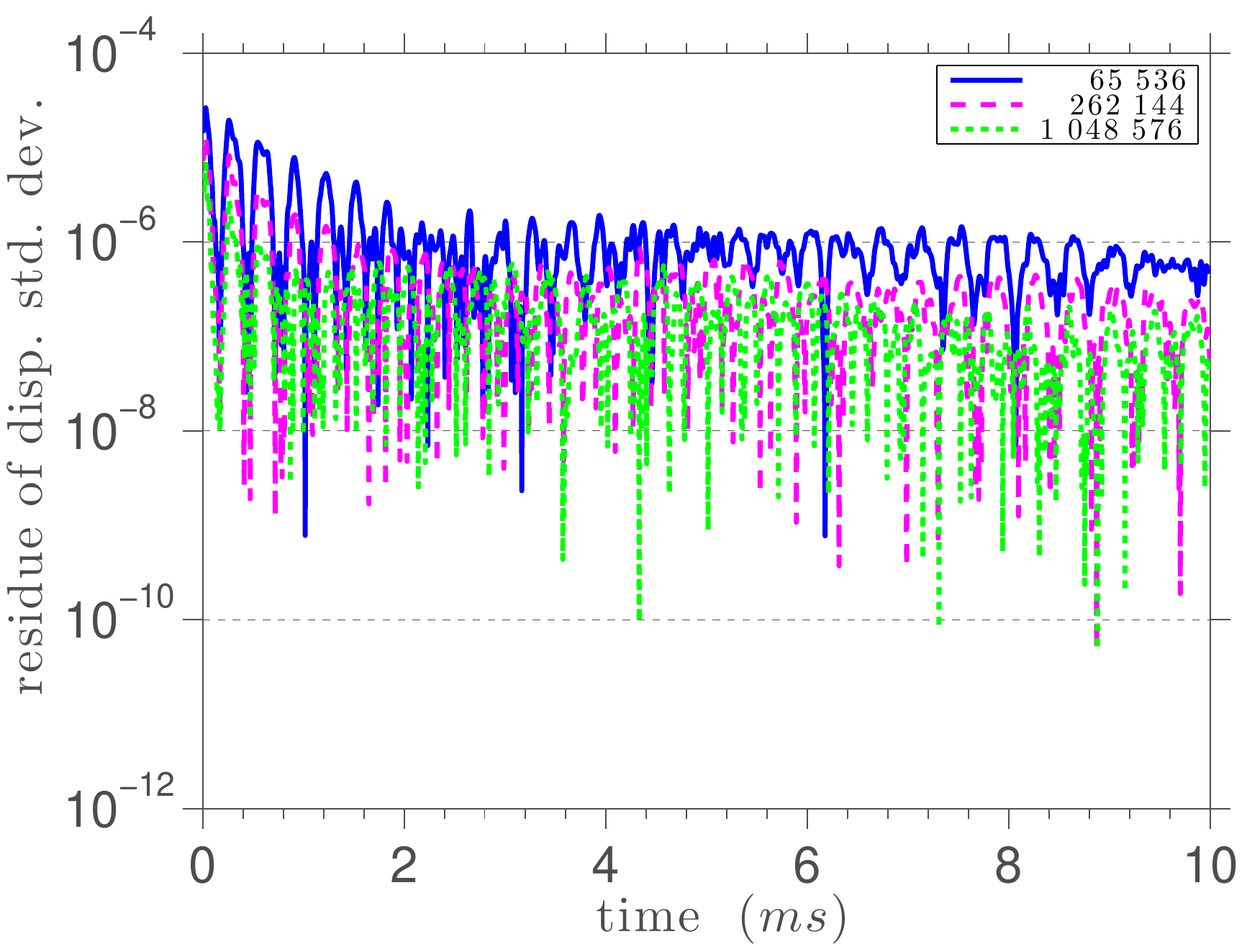}
				\caption{This figure illustrates the evolution of the residue of the 
							$\randproc{U}(L,\cdot,\cdot)$ standard deviation.}
				\label{res_std_uL_fig}
\end{figure}

Figure~\ref{res_mean_uL_fig} illustrates the evolution of several residues
of the $\randproc{U}(L,\cdot,\cdot)$ mean, and Figure~\ref{res_std_uL_fig}
illustrates the evolution of several residues of the $\randproc{U}(L,\cdot,\cdot)$ 
standard deviation. We note that, the logarithm of the mean value residues 
are almost always less than $\bigO{10^{-6}}$, for the case of statistics with 
larger samples, and presents an alternate behavior between large drops and 
climbs, (as seen in Figure~\ref{res_mean_uL_fig}). On the other hand, 
the logarithm of the standard deviation residue is greater than 
$\bigO{10^{-6}}$ in the initial instants. This behavior is not maintained after 
$2~ms$, when the residue curves keep their alternate behavior, but almost 
always below $\bigO{10^{-6}}$, as shown in Figure~\ref{res_std_uL_fig}.
These results show that statistics of first and second order may be obtained with 
great accuracy using the methodology presented in this work.

\subsection{Costs analysis for the processing time}

In what follows we will present an evaluation of the time spent 
by MC simulations based on the number of realizations. 
All the simulations were performed only once, without discarding 
any samples. Also, 20 virtual machines were used for the experiment, 
one for control and the other 19 for processing. Each task runned
uses $N_{serial}=256$.

\begin{table}[h!]
\begin{center}
	\caption{Comparison between the computational time spent (using 19 VMs) 
	by each one of the MC simulations in the cloud, and the corresponding speed-up 
	factors compared to a serial simulation.}
	\vspace{5mm}
	
%

\vspace{1mm}
\begin{tabular}{rrrrrrrc}
\toprule
 $N_{MC}$ & tasks & split & process & merge & total & serial & speed-up \\
          &       &  (ms) & (ms)    &  (ms) & (min) & (min) &           \\
\cmidrule{1-8}
	     256  &    1 &    --- &      111~998 &   2~250 &    1.9 &     1.9 &  1.0 \\
	  16~384  &   64 &    500 &      451~397 &  11~875 &    7.7 &   127.3 & 16.5 \\
	  65~536  &  256 &  2~328 &    1~576~711 &  27~031 &   26.7 &   509.3 & 19.0 \\
	  262~144 & 1024 &  7~609 &    6~078~757 &  89~450 &  102.9 & 2~037.0 & 19.8 \\
	1~048~576 & 4096 & 37~766 &   24~422~238 & 336~799 &  413.3 & 8~148.1 & 19.7 \\
\bottomrule
\end{tabular}

	\label{cpu_time_tab}
\end{center}
\end{table}

A comparison between the computational time spent by each one of 
the MC simulations in the cloud, and the corresponding 
speed-up factors compared with a serial simulation can be seen
in the Table~\ref{cpu_time_tab}. In this table the column $N_{MC}$ 
represents the total number of realizations of the experiment; 
the column \emph{tasks} represents the number of tasks to be processed; 
the columns \emph{split}, \emph{process} and \emph{merge} represent 
the time, in milliseconds, consumed in each stage of our parallelization strategy; 
the column \emph{total} represents, in minutes, the total time spent by the 
parallel MC simulation (split + process + merge); 
the column \emph{serial} shows, in minutes, the processing time of a 
MC simulation for the same problem executed in serial; 
and the column \emph{speed-up} shows a metric of the parallel simulation 
performance, defined as the ratio between the serial and the total time of simulation.

Before discussing the analysis of the results, we should indicate that
the serial time, shown in the first line of the Table~\ref{cpu_time_tab}, 
was obtained by extrapolation of the average processing time spent to 
run a single task in two virtual machines of the cloud. To obtain this average 
time the task was repeated 10 times (5 times in each VM).

The first thing that we can note in the Table~\ref{cpu_time_tab} is that 
the computational cost of the split and merge tasks are almost negligible 
when compared to the time spent by the process. Hence the importance 
of adopting a strategy of parallelism for the processing step. Also, we can 
see that, in all of the experiments, the strategy of parallelism in the cloud 
provided performance gains, and such gains are greater as the number of 
realizations increases. This shows the efficiency of our parallelization strategy 
in the case of study analyzed. Moreover, the speed-up observed is close to the 
number of virtual machines used, which indicates that the implementation of
the McCloud setting is good.

\begin{figure} [h!]
				\centering
				\includegraphics[scale=0.7]{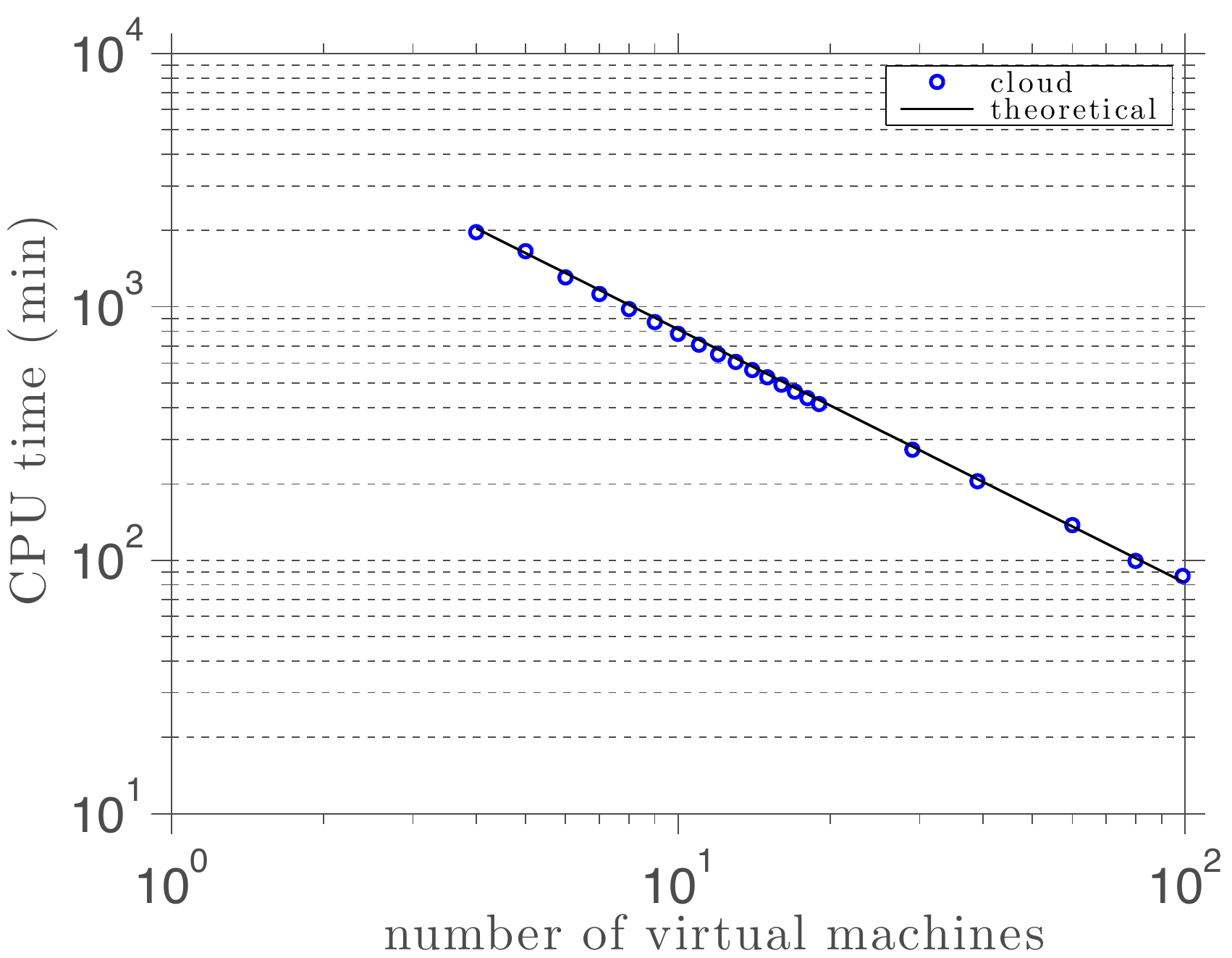}
				\caption{CPU time of a MC simulation, with 1 048 576 realizations,
				as function of the numbers of virtual machines.}
				\label{cpu_time_vs_vm}
\end{figure}

In order to evaluate the efficiency of the parallelization done in the cloud,
we studied how the processing time of the MC simulation, with $1,048,576$ 
realizations, varies as a function of the number of virtual machines used to 
perform the task. As can be seen in the Table~\ref{cpu_time2_tab}, the 
processing time decays monotonically as the number of VMs increases. 
This decreasing behavior is almost linear in logarithmic scale, as shown 
in the Figure~\ref{cpu_time_vs_vm}, and indicates the effectiveness of the 
parallelization of MC tasks in the cloud.

\begin{table}[h!]
\begin{center}
	\caption{Comparison between the computational time spent (using 19 VMs) 
	by each one of the MC simulations in the cloud, and the corresponding speed-up 
	factors compared to a serial simulation.}
	\vspace{5mm}
	
%

\vspace{1mm}
\begin{tabular}{rrrrrcr}
\toprule
   VMs & split & process & merge & total  & speed-up & cost \\
       &  (ms) & (ms)    &  (ms) & (min)  &          & (US\$)   \\
\cmidrule{1-7}
 4 & 37~734 & 117~637~861 &	431~691 & 1968.5 & 4.0 & 17.15 \\
 5 & 34~062 &  99~041~631 &	354~791 & 1657.2 & 4.7 & 18.23 \\
 6 & 37~923 &  78~040~047 &	343~105 & 1307.0 & 6.0 & 17.39 \\
 7 & 36~982 &  66~989~658 &	348~603 & 1122.9 & 7.0 & 17.63 \\
 8 & 27~390 &  58~431~244 &	349~099 &  980.1 & 8.0 & 18.11 \\
 9 & 29~407 &  51~818~238 &	343~720 &  869.9 & 9.0 & 18.11 \\
10 & 28~015 &  46~564~270 &	332~119 &  782.1 & 10.0 & 18.83 \\
11 & 28~734 &  42~192~535 &	334~432 &  709.3 & 11.0 & 17.99 \\
12 & 27~687 &  38~611~499 &	341~308 &  649.7 & 12.0 & 18.11 \\
13 & 34~874 &  36~039~896 &	334~806 &  606.8 & 12.9 & 19.55 \\
14 & 36~359 &  33~412~517 &	338~696 &  563.1 & 13.9 & 19.31 \\
15 & 36~499 &  31~199~806 &	343~874 &  526.3 & 14.8 & 18.83 \\
16 & 38~576 &  29~215~641 &	338~014 &  493.2 & 15.8 & 20.03 \\
17 & 40~484 &  27~406~752 &	340~522 &  463.1 & 16.8 & 19.19 \\
18 & 39~015 &  25~800~653 &	335~214 &  436.3 & 17.9 & 20.27 \\
19 & 37~766 &  24~422~238 &	336~799 &  413.3 & 18.9 & 19.07 \\
29 & 29~282 &  16~000~466 &	343~036 &  272.9 & 28.6 & 21.71 \\
39 & 29~547 &  11~927~263 &	346~594 &  205.1 & 38.0 & 24.23 \\
60 & 31~547 &   7~891~904 &	327~392 &  137.5 & 56.7 & 29.63 \\
80 & 25~328 &   5~900~347 &	 33~709 &   99.3 & 78.5 & 29.63 \\
99 & 29~578 &   4~834~623 &	340~015 &   86.7 & 89.9 & 36.47 \\
\bottomrule
\end{tabular}

	\label{cpu_time2_tab}
\end{center}
\end{table}

The small deviations from the theoretical curve of parallelization,
that can be seen in the Figure~\ref{cpu_time_vs_vm}, are due to the
management of tasks by the cloud. However, in the range analyzed,
between 4 and 99 virtual machines, these performance losses are 
negligible and do not compromise the efficiency of parallelization 
strategy.

\begin{figure} [h!]
				\centering
				\includegraphics[scale=0.7]{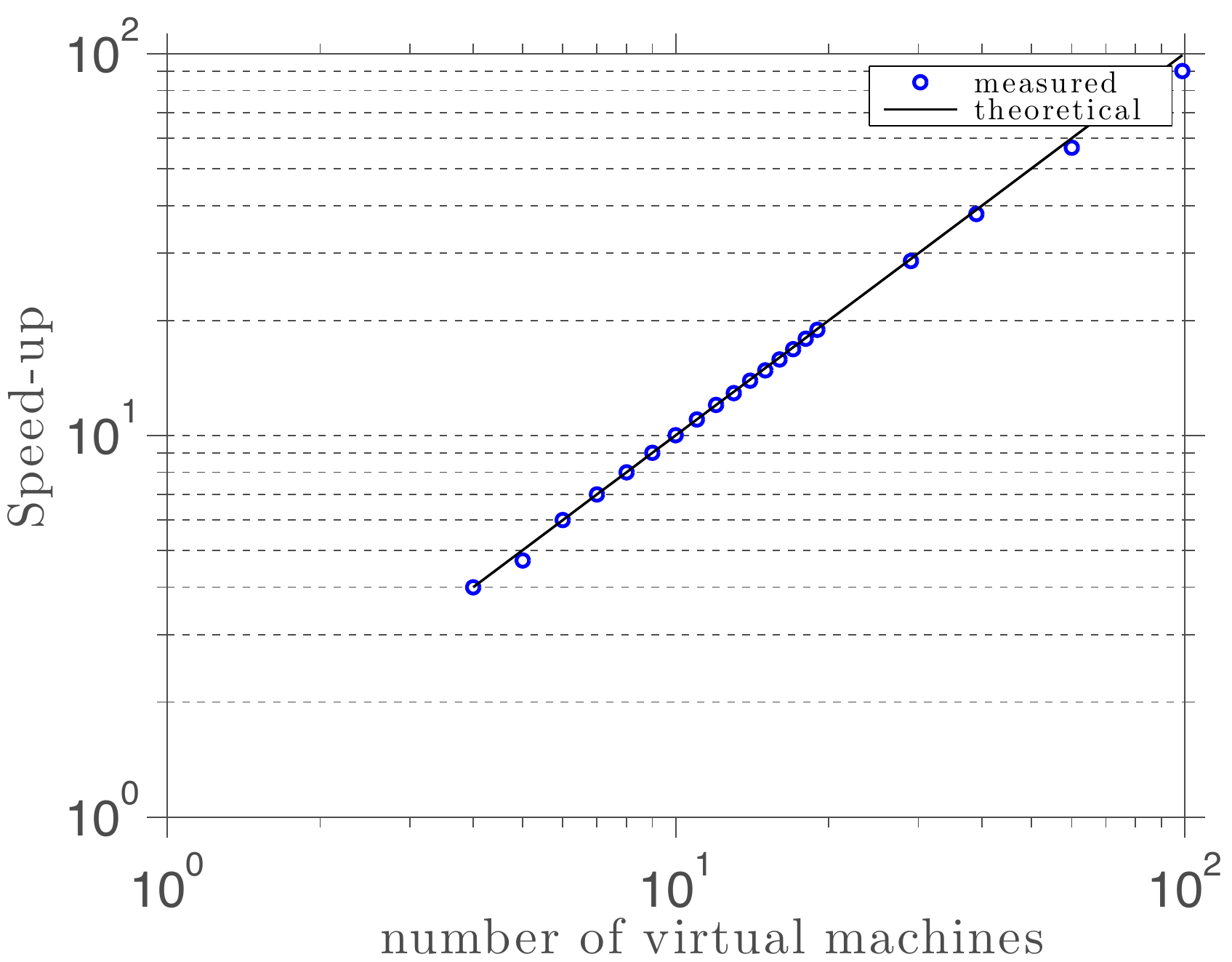}
				\caption{Speed-up of a MC simulation, with 1 048 576 realizations,
				as function of the numbers of virtual machines.}
				\label{speedup_vs_vm}
\end{figure}

Table~\ref{cpu_time2_tab} also shows that the speed-up increases
monotonically as the number of VMs grows. This behavior can be better 
appreciated in the Figure~\ref{speedup_vs_vm}, where the reader can verify 
that the values obtained for the speed-up are very close to the theoretical 
values, that would be obtained if there were no delays due to cloud 
management. The reader can note, also, that the measured values The ​corresponding 
to larger numbers of VMs (60, 80 and 99) are farther away from the theoretical 
reference. This occurs by the accumulation of management delays, which are 
larger for a higher amount of VMs.

\subsection{Costs analysis for the storage space}

A comparison between the storage space used by each one of the MC 
simulations in the cloud, and the financial cost associated with each simulation
can be seen in Table~\ref{space_cost_tab}. In this table the column
\emph{space} represents the total storage space, in $MB$, temporarily used
by the simulation, and the column \emph{cost} represents the cost of this 
experiment in US dollars.

\begin{table}[h!]
\begin{center}
	\caption{Comparison between the storage space used by each one of the MC 
	simulations in the cloud, and the financial cost associated with each simulation.}
	\vspace{5mm}
	
%
%
%

\vspace{5mm}
\begin{tabular}{rrr}
\toprule
 $N_{MC}$ & space & cost \\
                 & ($MB$) & ($US\$$) \\
\cmidrule(r){1-3}
	       16~384 &  12.4 &  5.39 \\
	       65~536 &  49.0 &  5.39 \\
	      262~144 & 195.2 &  7.67 \\
	    1~048~576 & 780.8 & 19.07 \\
\bottomrule
\end{tabular}

	\label{space_cost_tab}
\end{center}
\end{table}

Observing the data in Table~\ref{space_cost_tab} we can see that our 
parallelization strategy used fairly little storage space, even for a large number of 
MC realizations. Additionally, the financial cost, even for the most complex simulation,
is very small; which is a major advantage when compared to the costs of acquiring and 
maintaining a traditional cluster.


\section{Concluding remarks}
\label{concl_remarks}

We present a strategy for parallelizing the Monte Carlo method in the 
context of cloud computing, using the fundamental idea of the MapReduce 
paradigm. This strategy is described in detail and illustrated in the simulation 
of a simple problem of stochastic structural dynamics. The simulation results 
show good accuracy for low-order statistics, low storage space usage, and 
that the performance gains increase with the number of Monte Carlo realizations. 
It was also illustrated that even a simple problem can require many realizations for 
the convergence of histograms, which makes the cloud computing strategy very attractive, 
due to its high scalability capacity and low-cost. Thus, this article demonstrates 
that the advent of cloud computing can become an important enabler for the 
adoption of the MC method to compute the propagation of uncertainty in 
complex stochastic models.

\section*{Acknowledgments}

The authors are indebted to the Brazilian agencies CNPq, CAPES, and FAPERJ 
for the financial support given to this research. By the free use of the 
Windows Azure platform, the authors are very grateful to Microsoft Corporation.
Also, they wish to thank the anonymous referee, for useful comments and 
suggestions.

\bibliographystyle{elsarticle-num}


\end{document}